\documentclass{article}

\begin{document}

\title{Propagation of light in non-inertial reference frames}
\author{Vesselin Petkov \\
Science College, Concordia University\\
1455 De Maisonneuve Boulevard West\\
Montreal, Quebec, Canada H3G 1M8\\
vpetkov@alcor.concordia.ca}
\date{14 December 2003}
\maketitle

\begin{abstract}
It is shown that the complete description of the propagation of
light in a gravitational field and in non-inertial reference
frames in general requires an average coordinate and an average
proper velocity of light. The need for an average coordinate
velocity of light in non-inertial frames is demonstrated by
considering the propagation of two vertical light rays in the
Einstein elevator (in addition to the horizontal ray originally
discussed by Einstein). As an average proper velocity of light is
implicitly used in the Shapiro time delay (as shown in the
Appendix) it is explicitly derived and it is shown that for a
round trip of a light signal between two points in a gravitational
field the Shapiro time delay not only depends on which point it is
measured at, but in the case of a parallel gravitational field it
is not always a delay effect. The propagation of light in rotating
frames (the Sagnac effect) is also discussed and an expression for
the coordinate velocity of light is derived. The use of this
coordinate velocity naturally explains why an observer on a
rotating disk finds that two light signals emitted from a point on
the rim of the disk and propagating in opposite directions along
the rim do not arrive simultaneously at the same point.
\end{abstract}

\section{Introduction}

One of the fundamental facts of modern physics is the constancy of the speed
of light. Einstein regarded it as one of the two postulates on which special
relativity is based. So far, however, little attention has been paid to the
status of this postulate when teaching special relativity. It turns out that
the constancy of the speed of light is a direct consequence of the
relativity principle, not an independent postulate. To see this let us
consider the two postulates of special relativity as formulated by Einstein
in his 1905 paper "On the electrodynamics of moving bodies": "the same laws
of electrodynamics and optics will be valid for all frames of reference for
which the equations of mechanics hold good. We will raise this conjecture
(the purport of which will hereafter be called the "Principle of
Relativity") to the status of a postulate, and also introduce another
postulate, which is only apparently irreconcilable with the former, namely,
that light is always propagated in empty space with a definite velocity
\textit{c} which is independent of the state of the motion of the emitting
body" \cite{einstein}. As the principle of relativity states that "the laws
of physics are the same in all inertial reference frames" and the constancy
of the speed of light means that "the speed of light is the same in all
inertial reference frames (regardless of the motion of the source or the
observer)" it follow that the second postulate is indeed a consequence of
the first - the law describing the propagation of light is the same for all
inertial observers.

This becomes even clearer if it is taken into account that the
relativity principle is a statement of the impossibility to detect
absolute motion. Since all inertial observers (moving with
constant velocity) are completely equivalent (none is in absolute
motion) according to the principle of relativity, they all observe
the same phenomena and describe them by using the same laws of
physics. Therefore it does follow that light should propagate with
the same speed in all inertial frames; otherwise, if an inertial
observer found that the speed of light were not \textit{c }in his
reference frame, that observer would say that he detected his
absolute motion.

That all inertial observers are equivalent is also seen from the fact that
they are represented by \textit{geodesic} worldlines which in the case of
flat spacetime are \textit{straight} worldlines. However, when an observer
is accelerating his worldline is not geodesic (not a straight worldline in
flat spacetime). Therefore, accelerated motion, unlike motion with constant
velocity, is absolute - there is an absolute difference between a geodesic
and a non-geodesic worldline. This means that the laws of physics in
inertial and non-inertial reference frames are not the same. An immediate
consequence is that the speed of light is not constant in non-inertial
frames - a non-inertial observer can detect his accelerated motion by using
light signals.

It is precisely this corollary of special relativity that received little
attention in the courses and books on relativity. In fact, that corollary is
regularly used but since it is done implicitly confusions are not always
avoided. For instance, an observer in Einstein's thought experiment \cite%
{infeld} involving an accelerating elevator can discover his
accelerated motion by the deflection of a light ray from its
horizontal path. An observer in a rotating reference frame, a
rotating disk for example, can detect the disk accelerated motion
also by using light: light signals emitted from a point \textit{M}
in opposite directions along the rim of the disk do not arrive at
the same time at \textit{M} (this is the so called Sagnac effect)
\cite{rotating}. It is explicitly stated in general relativity
that the \textit{local} speed of light is always \textit{c}, which
implies that the speed of light along a \textit{finite} distance
is not necessarily \textit{c}. However, up to now no average
velocity of light propagating between two points has been defined.
Before introducing such a velocity let us consider several
examples that demonstrate the need for it.

Einstein's thought experiment \cite{infeld} involving an elevator at rest in
a \textit{parallel} \cite{parallel} gravitational field of strength $\mathbf{%
g}$ and an elevator accelerating with an acceleration $\mathbf{a=-g}$ was
designed to demonstrate the equivalence of the non-inertial reference frames
$N^{g}$ (elevator at rest in the gravitational field) and $N^{a}$ (elevator
accelerating in space devoid of gravity). Einstein called this equivalence
the principle of equivalence: it is not possible by experiment to
distinguish between the non-inertial frames $N^{a}$ and $N^{g}$ which means
that all physical phenomena look the same in $N^{a}$ and $N^{g}$. Therefore
if a horizontal light ray propagating in $N^{a}$ bends, a horizontal light
ray propagating in $N^{g}$ should bend as well.

Although even introductory physics textbooks \cite{tipler}-\cite{Fishbane}
have started to discuss Einstein's elevator experiment an obvious question
has been overlooked: \textquotedblright Are light rays propagating in an
elevator in a vertical direction (parallel and anti-parallel to $\mathbf{a}$
or $\mathbf{g}$) also affected by the accelerated motion of the elevator or
its being in a gravitational field?\textquotedblright\ The answer to this
question requires the introduction of an \textit{average }\emph{coordinate}
velocity of light which turns out to be different from $c$ in the case of
vertical light rays (see Figure~1 and the detailed discussion in Section 2).
It should be stressed that it is the average coordinate velocity of light
\textit{between two points} that is different from $c$; the local speed of
light measured at a point is always $c$.

A second average velocity of light - an \textit{average }\emph{proper}
velocity of light - is required for the explanation of the Shapiro time
delay \cite{shapiro64}, \cite{shapiro66}. It also turns out not to be always
$c$. The fact that it takes more time for a light signal to travel between
two points $P$ and $Q$ in a gravitational field than between the same points
in flat spacetime as determined by an observer at one of the points
indicates that the average velocity of light between the two points is
smaller than $c$. As the proper time of the observer is used in measuring
that velocity it seems appropriate to call it average \emph{proper} velocity
of light. Unlike the average coordinate velocity the average proper velocity
of light between two points depends on which point it is measured at. This
fact confirms the dependence of the Shapiro time delay on the point where it
is measured and shows, as we shall see in Section 3, that in the case of a
parallel gravitational field it is not always a delay effect (in such a
field the average proper velocity of light is defined in terms of both the
proper distance and proper time of an observer). A light signal will be
delayed \emph{only} if it is measured at a point $P$ that is farther from
the gravitating mass producing the parallel field; if it is measured at the
other point $Q,$ closer to the mass, it will take less time for the signal
to travel the same distance which shows that the average proper velocity of
the signal determined at $Q$ is greater than that measured at $P$ and
greater than $c$.

Due to the calculation of the average velocities of light in $N^{a}$ and $%
N^{g}$ to verify their agreement with the equivalence principle only a
parallel gravitational field will be considered in Sections 2 and 3. That is
why the expressions for the average light velocities will be derived for
this case. Their generalized expressions for the case of the Schwarzschild
metric and other metrics can be easily obtained (see Appendix).

Section 4 deals with the propagation of light in a rotating frame. In the
case of light signals propagating in opposite directions along the rim of a
rotating disk only the introduction of a coordinate velocity (depending on
the centripetal acceleration of the disk) can explain why an observer on the
disk discovers that the light signals do not arrive at the same time at the
source point.

The introduction of the average velocities of light also sheds some light on
a subtle feature of the propagation of light in the vicinity of a massive
body - whether or not light falls in its gravitational field. The particle
aspect of light seems to entail that a photon, like any other particle,
should fall in a gravitational field (due to the mass corresponding to its
energy); the deflection of light by a massive body appears to support such a
view. And indeed this view is sometimes implicitly or explicitly expressed
in papers and books although the correct explanation is given in some books
on general relativity (see for instance \cite{taylor}-\cite{rindler2001}).
It has been recently claimed that the issue of whether or not a charge
falling in a gravitational field radiates can be resolved by assuming that
the charge's electromagnetic field is also falling \cite{harpaz}. An
electromagnetic field falling in a gravitational field, however, implies
that light falls in the gravitational field as well. Even Einstein and
Infeld appear to suggest that as a light beam has mass on account of its
energy it will fall in a gravitational field: \textquotedblright A beam of
light will bend in a gravitational field exactly as a body would if thrown
horizontally with a velocity equal to that of light\textquotedblright\ \cite%
{infeld}. This comparison is not quite precise since the vertical component
of the velocity of the body will increase as it falls whereas the velocity
of the \textquotedblright falling\textquotedblright\ light beam is
decreasing for a non-inertial observer (supported in a gravitational field)
, as we shall see below. Statements such as \textquotedblright a beam of
light will accelerate in a gravitational field, just like objects that have
mass\textquotedblright\ and therefore \textquotedblright near the surface of
the earth, light will fall with an acceleration of 9.81 $m/s^{2}$%
\textquotedblright\ have found their way in introductory physics textbooks
\cite{tipler}. It will be shown in Section 3 that during its
\textquotedblright fall\textquotedblright\ in a gravitational field light is
slowing down - a negative acceleration of 9.81 $m/s^{2}$ is decreasing its
velocity (at a point $P$ near the Earth's surface as seen from another point
above or below $P$).

\section{Average coordinate velocity of light}

Why the average velocity of light between two points in a gravitational
field is not generally equal to $c$ can be most clearly shown by considering
two extra light rays parallel and anti-parallel to the gravitational
acceleration $\mathbf{g}$ in the Einstein thought experiment involving an
elevator at rest in the Earth's gravitational field (Figure~1).

\begin{center}
\begin {picture}(0,100)(-33,130)
\setlength{\unitlength}{0.25mm}
\put(-100,100){\framebox(110,220){}}
\put(10,212){\circle*{4}}
\put(-6,217){$B$}
\put(29,212){\line(1,0){10}}
\put(34,201){\vector(0,1){11}}
\put(10,190){\circle*{4}}
\put(-6,175){$B^{\prime}$}
\put(29,190){\line(1,0){10}}
\put(34,201){\vector(0,-1){11}}
\put(40,198){$\delta =\frac{1}{2} gt^{2}=gr^{2}/2c^{2}$}
\put(-5,305){$A$}
\put(10,320){\circle*{4}}
\put(10,100){\circle*{4}}
\put(-5,105){$C$}
\put(-100.5,212){\circle*{4}}
\put(-95,217){$D$}

\qbezier(-100.5,212)(-20,212)(2.5,193)
\put(2.5,193.3){\vector(3,-2){2}}
\put(-121,96.6){---}
\put(-121,317.1){---}
\put(-114,200){\vector(0,1){119.5}}
\put(-114,220.1){\vector(0,-1){119}}
\put(-131,212){$2r$}
\put(-100,80){\line(0,1){10}}
\put(10,80){\line(0,1){10}}
\put(-45,85){\vector(1,0){53.7}}
\put(-45,85){\vector(-1,0){53.7}}
\put(-50,87.5){$r$}
\put(17,320.5){\vector(0,-1){129}}
\put(17,99){\vector(0,1){89}}
\put(-144,84){\vector(0,1){249}}
\put(-157,212){$z$}
\thicklines
\put(-45,281){\vector(0,-1){33}}
\put(-39,267){{\bf g}}
\end {picture}
\end{center}

\vspace{2cm}

\begin{list}{}{\leftmargin=0em \rightmargin=0em}\item[]
\begin{center}
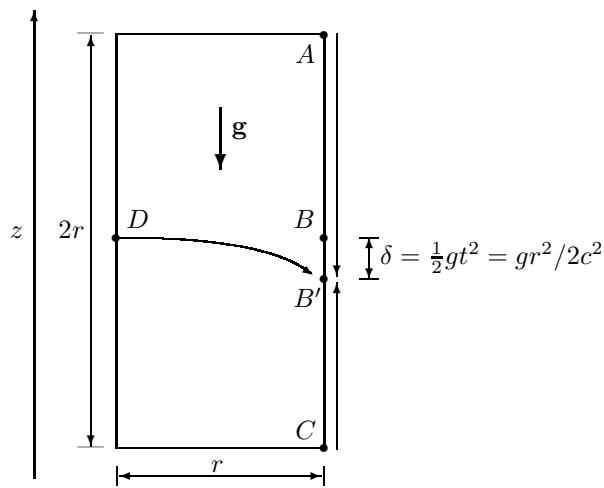
{\small Figure 1 - Propagation of light in the Einstein elevator at rest in a parallel gravitational field.}
\end{center}
\end{list}\vspace{0.1in}


Three light rays are emitted simultaneously in the elevator which is at rest
in the non-inertial reference frame $N^{g}$. Two rays are emitted from
points $A$ and $C$ towards point $B$ and the third light ray is following
the null path from $D$, spatially directed along constant $z$ towards $B$,
to $B^{\prime }$. The emission of the three rays is also simultaneous in the
local Lorentz (inertial) frame $I$ which is momentarily at rest with respect
to $N^{g}$ at the moment the light rays are emitted ($I$ and $N^{g}$ have a
common instantaneous three-dimensional space at this moment and therefore
common simultaneity). At the next moment, as $I$ starts to fall in the
gravitational field, it will appear to an observer in $I$ that the elevator
moves upwards with an acceleration $g=\left\vert \mathbf{g}\right\vert $.
Therefore as seen from $I$ the three light rays will arrive simultaneously
not at point $B$, but at $B^{\prime }$ since for the time $t=r/c$ the
elevator moves (from $I$'s viewpoint) at a distance $\delta
=gt^{2}/2=gr^{2}/2c^{2}$. As the simultaneous arrival of the three rays at
point $B^{\prime }$ is an \emph{absolute} (observer-independent) fact due to
its being a \emph{single} event, it follows that the rays arrive
simultaneously at $B^{\prime }$ as seen from $N^{g}$ as well. Since for the
\emph{same} coordinate time $t=r/c$ in $N^{g}$ the three light rays travel
different distances $DB^{\prime }\approx r$, $AB^{\prime }=r+\delta $, and $%
CB^{\prime }=r-\delta $ before arriving simultaneously at $B^{\prime }$ an
observer in $N^{g}$ concludes (to within terms $\sim c^{-2}$ \cite{g}) that
the \emph{average} velocity of the light ray propagating from $A$ to $%
B^{\prime }$ is slightly greater than $c$

\[
{c}_{AB^{\prime}}^{g}=\frac{r+\delta}{t} \approx c\left( 1+\frac{gr}{2c^{2}}%
\right).
\]

\noindent The average velocity ${c}_{CB^{\prime }}^{g}$ of the light ray
propagating from $C$ to $B^{\prime }$ is slightly smaller than $c$

\[
{c}_{CB^{\prime }}^{g}=\frac{r-\delta }{t}\approx c\left( 1-\frac{gr}{2c^{2}}%
\right) .
\]%
It is easily seen that to within terms $\sim c^{-2}$ the average velocity of
light between $A$ and $B$ is equal to that between $A$ and $B^{\prime }$,
i.e. $c_{AB}^{g}=c_{AB^{\prime }}^{g}$ and also $c_{CB}^{g}=c_{CB^{\prime
}}^{g}$:
\begin{equation}
c_{AB}^{g}=\frac{r}{t-\delta /c}=\frac{r}{t-gt^{2}/2c}=\frac{c}{1-gr/2c^{2}}%
\approx c\left( 1+\frac{gr}{2c^{2}}\right)  \label{c_g-AB}
\end{equation}%
and
\begin{equation}
c_{CB}^{g}=\frac{r}{t+\delta /c}\approx c\left( 1-\frac{gr}{2c^{2}}\right) .
\label{c_g-CB}
\end{equation}%
As the average velocities (\ref{c_g-AB}) and (\ref{c_g-CB}) are not
determined with respect to a specific point since the \emph{coordinate} time
$t$ is involved in their calculation, it is clear that (\ref{c_g-AB}) and (%
\ref{c_g-CB}) represent the average \emph{coordinate} velocities of light
between the points $A$ and $B$ and $C$ and $B$, respectively.

These expressions for the average coordinate velocity of light in $N^{g}$
can be also obtained from the coordinate velocity of light at a point in a
parallel gravitational field. In such a field proper and coordinate times do
not coincide whereas proper and coordinate distances are the same \cite%
{rindler68} as follows from the standard spacetime interval in a \emph{%
parallel} gravitational field \cite{misner}

\[
ds^{2}=\left( 1+\frac{2gz}{c^{2}}\right) c^{2}dt^{2}-dx^{2}-dy^{2}-dz^{2}
\]
which can be also written as \cite[p. 173]{misner}

\begin{equation}
ds^{2}=\left( 1+\frac{gz}{c^{2}}\right) ^{2}c^{2}dt^{2}-dx^{2}-dy^{2}-dz^{2}.
\label{ds_g}
\end{equation}%
Note that due to the existence of a horizon at $z=-c^{2}/g$ \cite[pp. 169,
172-173]{misner} there are constraints on the size of non-inertial reference
frames (accelerated or at rest in a parallel gravitational field) which are
represented by the metric (\ref{ds_g}). If the origin of $N^{g}$ is changed,
say to $z_{B}=0$ (See Figure~1), the horizon moves to $z=-c^{2}/g-\left\vert
z_{B}\right\vert $.

The coordinate velocity of light at a point $z$ can be obtained from (\ref%
{ds_g}) (for $ds^{2}=0$)
\begin{equation}
c^{g}\left( z\right) \equiv \frac{dz}{dt}=\pm c\sqrt{\left( 1+\frac{gz}{c^{2}%
}\right) ^{2}}=\pm \ c\left( 1+\frac{gz}{c^{2}}\right) .  \label{c_g_coord}
\end{equation}%
The $+$ and $-$ signs are for light propagating along or against $+\ z$,
respectively. Therefore, the coordinate velocity of light at a point $z$ is
locally isotropic in the $z$ direction. It is clear that the coordinate
velocity (\ref{c_g_coord}) cannot become negative due to the constraints on
the size of non-inertial frames which ensure that $\left\vert z\right\vert
<c^{2}/g$ \cite[pp. 169, 172]{misner}.

As seen from (\ref{c_g_coord}) the coordinate velocity of light is a
function of $z$ which shows that we can calculate the average coordinate
velocity between $A$ and $B$ by taking an average over the distance from $A$
to $B$. As $c^{g}\left( z\right) $ is not only continuous on the interval $%
\left[ z_{A},\ z_{B}\right] $ (for $\left\vert z\right\vert <c^{2}/g$), but
is also a linear function of $z$, we can write

\begin{equation}
c_{AB}^{g}=\frac{1}{z_{B}-z_{A}}\int_{z_{A}}^{z_{B}}c^{g}\left( z\right)
dz=c\left( 1+\frac{gz_{B}}{c^{2}}+\frac{gr}{2c^{2}}\right) ,
\label{c_g_av_z}
\end{equation}
where we took into account that $z_{A}=z_{B}+r$. When the coordinate origin
is at point $B$ ($z_{B}=0$) the expression (\ref{c_g_av_z}) coincides with (%
\ref{c_g-AB}).

The coordinate velocity of light $c^{g}\left( z\right) $ is also continuous
on the interval $\left[ t_{A},\ t_{B}\right] $, but in order to calculate $%
c_{AB}^{g}$ by taking an average of the velocity of light over the time of
its propagation from $A$ to $B$ we should find the dependence of $z$ on $t$.
From (\ref{ds_g}) we can write (for $ds^{2}=0$):

\[
dz=c\left( 1+\frac{gz}{c^{2}}\right) dt.
\]
By integrating and keeping only the terms proportional to $c^{-2}$ we find
that $z=ct$ which shows that $c^{g}\left( z\right) $ is also linear in $t$
(to within terms proportional to $c^{-2}$):

\[
c^{g}\left( t\right) =\pm \ c\left( 1+\frac{gt}{c}\right) .
\]

Therefore for the average coordinate velocity of light between points $A$
and $B$ we have:

\begin{eqnarray}
c_{AB}^{g} &=&\frac{1}{t_{B}-t_{A}}\int_{t_{A}}^{t_{B}}c^{g}\left( z\right)
dt=\frac{1}{t_{B}-t_{A}}\int_{t_{A}}^{t_{B}}c\left( 1+\frac{gz}{c^{2}}%
\right) dt  \nonumber \\
&=&\frac{1}{t_{B}-t_{A}}\int_{t_{A}}^{t_{B}}c\left( 1+\frac{gt}{c}\right)
dt=c\left( 1+\frac{gz_{B}}{c^{2}}+\frac{gr}{2c^{2}}\right) ,
\label{c_g_av_t}
\end{eqnarray}%
where the magnitude of $c^{g}\left( z\right) $ was used and it was
taken into account that $z_{A}=z_{B}+r$ and $z_{A}=ct_{A}$ and
$z_{B}=ct_{B}$. As expected this expression coincides with
(\ref{c_g_av_z}) and for $z_{B}=0$ is equal to (\ref{c_g-AB}).

The fact that $c^{g}\left( z\right) $ is linear in both $z$ and $t$ (to
within terms $\sim c^{-2}$) makes it possible to calculate the average
coordinate velocity of light propagating between $A$ and $B$ (See Figure~1)
by using the values of $c^{g}\left( z\right) $ only at the end points $A$
and $B$:
\[
c_{AB}^{g}=\frac{1}{2}\left( c_{A}^{g}+c_{B}^{g}\right) =\frac{1}{2}\left[
c\left( 1+\frac{gz_{A}}{c^{2}}\right) +c\left( 1+\frac{gz_{B}}{c^{2}}\right) %
\right]
\]%
and as $z_{A}=z_{B}+r$
\begin{equation}
c_{AB}^{g}=c\left( 1+\frac{gz_{B}}{c^{2}}+\frac{gr}{2c^{2}}\right) .
\label{c_g__AB}
\end{equation}%
This expression coincides with the expressions for $c_{AB}^{g}$ in (\ref%
{c_g_av_z}) and (\ref{c_g_av_t}).

For the average coordinate velocity of light propagating between $B$ and $C$
we obtain
\begin{equation}
c_{BC}^{g}=c\left( 1+\frac{gz_{B}}{c^{2}}-\frac{gr}{2c^{2}}\right)
\label{c_g_CB}
\end{equation}
since $z_{C}=z_{B}-r$. As noted above when the coordinate origin is at point
$B$ ($z_{B}=0$) the expressions (\ref{c_g__AB}) and (\ref{c_g_CB}) coincide
with (\ref{c_g-AB}) and (\ref{c_g-CB}).

The average coordinate velocities (\ref{c_g__AB}) and (\ref{c_g_CB})
correctly describe the propagation of light in $N^{g}$ yielding the right
expression $\delta =gr^{2}/2c^{2}$ (See Figure~1). It should be stressed
that without these average coordinate velocities the fact that the light
rays emitted from $A$ and $C$ arrive not at $B,$ but at $B^{\prime }$ cannot
be explained.

As a coordinate velocity, the average coordinate velocity of light is not
determined with respect to a specific point and depends on the choice of the
coordinate origin. Also, it is the same for light propagating from $A$ to $B$
and for light travelling in the opposite direction, i.e. $%
c_{AB}^{g}=c_{BA}^{g}$. Therefore, like the coordinate velocity (\ref%
{c_g_coord}) the average coordinate velocity is also isotropic. Notice,
however, that the average coordinate velocity of light is isotropic in a
sense that the average light velocity between two points is the same in both
directions. But as seen from (\ref{c_g__AB}) and (\ref{c_g_CB}) the average
coordinate velocity of light between different pairs of points, whose points
are the same distance apart, is different. As a result, as seen in Figure~1,
the light ray emitted at $A$ arrives at $B$ \textit{before} the light ray
emitted at $C$.

In an elevator (at rest in the non-inertial reference frame $N^{a}$)
accelerating with an acceleration $a=\left\vert \mathbf{a}\right\vert $,
where the metric is \cite[p. 173]{misner}

\begin{equation}
ds^{2}=\left( 1+\frac{az}{c^{2}}\right) ^{2}c^{2}dt^{2}-dx^{2}-dy^{2}-dz^{2},
\label{ds_a}
\end{equation}
the expressions for the average coordinate velocity of light between $A$ and
$B$ and $B$ and $C$, respectively, are

\begin{equation}
c_{AB}^{a}=c\left( 1+\frac{az_{B}}{c^{2}}+\frac{ar}{2c^{2}}\right)
\label{ca1}
\end{equation}%
and

\begin{equation}
c_{BC}^{a}=c\left( 1+\frac{az_{B}}{c^{2}}-\frac{ar}{2c^{2}}\right)
\label{ca2}
\end{equation}
in agreement with the equivalence principle.

\section{Average proper velocity of light}

The average coordinate velocity of light explains the propagation of light
in the Einstein elevator and in non-inertial reference frames in general,
but cannot be used in a situation where the average light velocity between
two points (say a source and an observation point) is determined \textit{%
with respect to one of the points}. Such situations occur in the Shapiro
time delay. As the local velocity of light is $c$ the average velocity of
light between a source and an observation point depends on which of the two
points is regarded as a reference point with respect to which the average
velocity is determined (at the reference point the local velocity of light
is always $c$). The dependence of the average velocity on which point is
chosen as a reference point demonstrates that that velocity is anisotropic.
This anisotropic velocity can be regarded as an average\emph{\ proper}
velocity of light since it is determined with respect to a given point and
therefore its calculation involves the proper time at that point. It is also
defined in terms of the proper distance as determined by an observer at the
same point (in the case of a parallel gravitational field).

Consider a light source at point $B$ (See Figure~1). To calculate the
average proper velocity of light originating from $B$ and observed at $A$
(that is, as seen from $A$) we have to determine the initial velocity of a
light signal at $B$ and its final velocity at $A$, both with respect to $A$.
As the local velocity of light is $c$ the final velocity of the light signal
determined at $A$ is obviously $c$. Noting that in a parallel gravitational
field proper and coordinate distances are the same \cite{rindler68} we can
determine the initial velocity of the light signal at $B$ as seen from $A$
\[
c_{B}^{g}=\frac{dz_{B}}{d\tau _{A}}=\frac{dz_{B}}{dt}\frac{dt}{d\tau _{A}}
\]%
where $dz_{B}/dt=c^{g}\left( z_{B}\right) $ is the coordinate velocity of
light (\ref{c_g_coord}) at $B$%
\[
c^{g}\left( z_{B}\right) =c\left( 1+\frac{gz_{B}}{c^{2}}\right)
\]%
and $d\tau _{A}=ds_{A}/c$ is the proper time for an observer with constant
spatial coordinates at $A$%
\[
d\tau _{A}=\left( 1+\frac{gz_{A}}{c^{2}}\right) dt.
\]%
As $z_{A}=z_{B}+r$ and $gz_{A}/c^{2}<1$ (since for any value of $z$ in $%
N^{g} $ there is a restriction $\left\vert z\right\vert <c^{2}/g$) for the
coordinate time $dt$ we have (to within terms $\sim c^{-2}$)
\[
dt\approx \left( 1-\frac{gz_{A}}{c^{2}}\right) d\tau _{A}=\left( 1-\frac{%
gz_{B}}{c^{2}}-\frac{gr}{c^{2}}\right) d\tau _{A}.
\]%
Then for the initial velocity $c_{B}^{g}$ at $B$ as seen from $A$ we obtain
\[
c_{B}^{g}=c\left( 1+\frac{gz_{B}}{c^{2}}\right) \left( 1-\frac{gz_{B}}{c^{2}}%
-\frac{gr}{c^{2}}\right)
\]%
or keeping only the terms $\sim c^{-2}$
\begin{equation}
c_{B}^{g}=c\left( 1-\frac{gr}{c^{2}}\right) .  \label{c-B_at-A}
\end{equation}%
Therefore an observer at $A$ will determine that a light signal is emitted
at $B$ with the velocity (\ref{c-B_at-A}) and during the time of its journey
towards $A$ (away from the Earth's surface) will \textit{accelerate} with an
acceleration $g$ and will arrive at $A$ with a velocity exactly equal to $c$.

For the average proper velocity $\bar{c}_{BA}^{g}=(1/2)(c_{B}^{g}+c)$ of
light propagating from $B$ to $A$ as seen from $A$ we have
\begin{equation}
\bar{c}_{BA}^{g}\left( as\ seen\ from\ A\right) =c\left( 1-\frac{gr}{2c^{2}}%
\right) .  \label{c-g_BA}
\end{equation}

As the local velocity of light at $A$ (measured at $A$) is $c$ it follows
that if a light signal propagates from $A$ towards $B$ its initial velocity
at $A$ is $c$, its final velocity at $B$ is (\ref{c-B_at-A}) and therefore,
as seen from $A$, it is subjected to a negative acceleration $g$ and will
\textit{slow down} as it \textquotedblright falls\textquotedblright\ in the
Earth's gravitational field. This shows that the average proper speed $\bar{c%
}_{AB}^{g}\left( as\ seen\ from\ A\right) $ of a light signal emitted at $A$
with the initial velocity $c$ and arriving at $B$ with the final velocity (%
\ref{c-B_at-A}) will be equal to the average proper speed $\bar{c}%
_{BA}^{g}\left( as\ seen\ from\ A\right) $ of a light signal propagating
from $B$ towards $A$. Thus, as seen from $A$, the back and forth average
proper speeds of light travelling between $A$ and $B$ are the \emph{same}.

Now let us determine the average proper velocity of light between $B$ and $A$
with respect to point $B$. A light signal emitted at $B$ as seen from $B$
will have an initial (local) velocity $c$ there. The final velocity of the
signal at $A$ as seen from $B$ will be
\[
c_{A}^{g}=\frac{dz_{A}}{d\tau _{B}}=\frac{dz_{A}}{dt}\frac{dt}{d\tau _{B}}
\]%
where $dz_{A}/dt=c^{g}\left( z_{A}\right) $ is the coordinate velocity of
light at $A$%
\[
c^{g}\left( z_{A}\right) =c\left( 1+\frac{gz_{A}}{c^{2}}\right)
\]%
and $d\tau _{B}$ is the proper time at $B$%
\[
d\tau _{B}=\left( 1+\frac{gz_{B}}{c^{2}}\right) dt.
\]%
Then as $z_{A}=z_{B}+r$ we obtain for the velocity of light at $A$ as
determined from\emph{\ }$B$
\begin{equation}
c_{A}^{g}=c\left( 1+\frac{gr}{c^{2}}\right) .  \label{c-A_at-B}
\end{equation}

Using (\ref{c-A_at-B}) the average proper velocity of light propagating from
$B$ to $A$ as determined from\emph{\ }$B$ becomes
\begin{equation}
\bar{c}_{BA}^{g}\left( as\ seen\ from\ B\right) =c\left( 1+\frac{gr}{2c^{2}}%
\right) .  \label{c-g-AB}
\end{equation}%
If a light signal propagates from $A$ to $B$ its average proper speed $\bar{c%
}_{AB}^{g}\left( as\ seen\ from\ B\right) $ will be equal to $\bar{c}%
_{BA}^{g}\left( as\ seen\ from\ B\right) $ - the average proper speed of
light propagating from $B$ to $A$. This demonstrates that for an observer at
$B$ a light signal emitted from $B$ with a velocity $c$ will \textit{%
accelerate} towards $A$ with an acceleration $g$ and will arrive there with
the final velocity (\ref{c-A_at-B}). As determined by the $B$-observer a
light signal emitted from $A$ with the initial velocity (\ref{c-A_at-B})
will be \textit{slowing down} (with $-\ g$) as it \textquotedblright
falls\textquotedblright\ in the Earth's gravitational field and will arrive
at $B$ with a final velocity exactly equal to $c$. Therefore an observer at $%
B$ will agree with an observer at $A$ that a light signal will \textit{%
accelerate} with an acceleration $g$ on its way from $B$ to $A$ and will
\textit{decelerate} while "falling" in the Earth's gravitational field
during its propagation from $A$ to $B$ but disagree on the velocity of light
at the points $A$ and $B$.

Comparing (\ref{c-g_BA}) and (\ref{c-g-AB}) demonstrates that the two
average proper speeds between the same points $A$ and $B$ are not equal and
depend on where they are measured from. As we expected the fact that the
local velocity of light at the reference point is $c$ makes the average
proper velocity between two points dependant on where the reference point
is. An immediate consequence from here is that the Shapiro time delay does
not always mean that it takes more time for light to travel a given distance
in a parallel gravitational field than the time needed in flat spacetime.

In the case of a parallel gravitational field the Shapiro time effect for a
round trip of a light signal propagating between $A$ and $B$ determined from
point $A$ will be indeed a delay effect:

\[
\Delta \tau _{A}=\frac{2r}{c\left( 1-gr/2c^{2}\right) }\approx \Delta
t_{flat}\left( 1+\frac{gr}{2c^{2}}\right) ,
\]%
where $\Delta t_{flat}=2r/c$ is the time for the round trip of light between
$A$ and $B$ in flat spacetime. However, an observer at $B$ will determine
that it takes less time for a light signal to complete the round trip
between $A$ and $B$:

\[
\Delta \tau _{B}=\frac{2r}{c\left( 1+gr/2c^{2}\right) }\approx \Delta
t_{flat}\left( 1-\frac{gr}{2c^{2}}\right) .
\]%
However, in the Schwarzschild metric the Shapiro effect is always
a delay effect since the average proper speed of light in that
metric is always smaller that \textit{c} as shown in the Appendix.

The average proper velocity of light between $A$ and $B$ can be also
obtained by using the average coordinate velocity of light (\ref{c_g__AB})
between the same points:

\[
c_{AB}^{g}\equiv \frac{r}{\Delta t}=c\left( 1+\frac{gz_{B}}{c^{2}}+\frac{gr}{%
2c^{2}}\right)
\]%
Let us calculate the average proper velocity of light propagating between $A$
and $B$ as determined from point $A.$ This means that we will use $A$'s
proper time $\Delta \tau _{A}=\left( 1+gz_{A}/c^{2}\right) \Delta t$:

\[
\bar{c}_{AB}^{g}(as\ seen\ from\ A)=\frac{r}{\Delta \tau _{A}}=\frac{r}{%
\Delta t}\frac{\Delta t}{\Delta \tau _{A}}
\]
Noting that $r/\Delta t$ is the average coordinate velocity (\ref{c_g__AB})
and $z_{A}=z_{B}+r$ we have (to within terms $\sim c^{-2}$)

\[
\bar{c}_{AB}^{g}(as\ seen\ from\ A)\approx c\left( 1+\frac{gz_{B}}{c^{2}}+%
\frac{gr}{2c^{2}}\right) \left( 1-\frac{gz_{A}}{c^{2}}\right) \approx
c\left( 1-\frac{gr}{2c^{2}}\right)
\]%
which coincides with (\ref{c-g_BA}).

The calculation of the average proper velocity of light propagating between $%
A$ and $B$, but as seen from $B$ yields the same expression as (\ref{c-g-AB}%
):

\[
\bar{c}_{AB}^{g}(as\ seen\ from\ B)=\frac{r}{\Delta \tau _{B}}=\frac{r}{%
\Delta t}\frac{\Delta t}{\Delta \tau _{B}}\approx c\left( 1+\frac{gz_{B}}{%
c^{2}}+\frac{gr}{2c^{2}}\right) \left( 1-\frac{gz_{B}}{c^{2}}\right) \approx
c\left( 1+\frac{gr}{2c^{2}}\right) .
\]

As evident from (\ref{c-g_BA}) and (\ref{c-g-AB}) the average proper
velocity of light emitted from a common source and determined at different
points around the source is anisotropic in $N^{g}$ - if the observation
point is above the light source the average proper speed of light is
slightly smaller than $c$\ and smaller than the average proper speed as
determined from an observation point below the source. If an observer at
point $B$ (See Figure~1) determines the average proper velocities of light
coming from $A$ and $C$ he will find that they are also anisotropic - the
average proper velocity of light coming from $A$ is greater than that
emitted at $C$ and therefore the light from $A$ will arrive at $B$ \textit{%
before} the light from $C$ (provided that the two light signals from $A$ and
$C$ are emitted simultaneously in $N^{g}$). However, if the observer at $B$
(See Figure~1) determines the back and forth average proper speeds of light
propagating between $A$ and $B$ he finds that they are the same (the back
and forth average proper speeds of light between $B$ and $C$ are also the
same).

The calculation of the average proper velocities of light in an accelerating
frame $N^{a}$ gives:
\[
c_{BA}^{a}\left( as\ seen\ from\ A\right) =c\left( 1-\frac{ar}{2c^{2}}%
\right)
\]%
and
\[
c_{BA}^{a}\left( as\ seen\ from\ B\right) =c\left( 1+\frac{ar}{2c^{2}}%
\right) .
\]%
where $a\mathbf{=|a|}$ is the frame's proper acceleration.

\section{The Sagnac effect}

The Sagnac effect can de described as follows. Two light signals
emitted from a point \textit{M} on the rim of a rotating disk and
propagating along its rim in opposite directions will not arrive
simultaneously at \textit{M}. There still exist people who
question special relativity and their main argument has been this
effect. They claim that for an observer on the rotating disk the
speed of light is not constant - that the Galilean law of velocity
addition ($c+v$ and $c-v$, where $v$ is the orbital speed at a
point on the disk rim) should be used by the rotating observer in
order to explain the time difference in the arrival of the two
light signals at \textit{M}. What makes such claims even more
persistent is the lack of a clear position on the issue of the
speed of light in non-inertial reference frames. What special
relativity states is that the speed of light is constant only in
inertial reference frames - this constancy follows from the
impossibility to detect absolute motion (more precisely, it
follows from the non-existence of absolute motion). Accelerated
motion can be detected and for this reason the coordinate velocity
of light in non-inertial reference frames is a function of the
proper acceleration of the frame. The rotating disk is a
non-inertial reference frame and its acceleration can be detected
by different means including light signals. That is why it is not
surprising that the coordinate velocity of light as determined on
the disk depends on the centripetal acceleration of the disk. As
we shall see below the coordinate velocity of light calculated on
the disk is \textit{not} a manifestation of the Galilean law of
velocity addition.

Consider two disks whose centers coincide. One of them is stationary, the
other rotates with constant angular velocity $\omega $. As the stationary
disk can be regarded as an inertial frame its metric is the Minkowski metric:

\begin{equation}
ds^{2}=c^{2}dt^{2}-dx^{2}-dy^{2}-dz^{2}.  \label{s1}
\end{equation}%
To write the interval $ds^{2}$ in polar coordinates we use the transformation

\begin{equation}
t=t\hspace{0.3cm}\hspace{0.5cm}x=R\cos \Phi \hspace{0.3cm}\hspace{0.5cm}%
y=R\sin \Phi \hspace{0.3cm}\hspace{0.5cm}z=z.  \label{s2}
\end{equation}%
By substituting (\ref{s1}) in (\ref{s2}) we get

\begin{equation}
ds^{2}=c^{2}dt^{2}-dR^{2}-R^{2}d\Phi ^{2}-dz^{2}.  \label{s3}
\end{equation}%
Let an observer on the rotating disk use the coordinates $t$, $r$, $\varphi $%
, and $z$. The transformation between the coordinates on the
stationary and on the rotating disk is obviously:

\begin{equation}
t=t\hspace{0.3cm}\hspace{0.5cm}R=r\hspace{0.3cm}\hspace{0.5cm}\Phi =\varphi
+\omega t\hspace{0.3cm}\hspace{0.5cm}z=z.  \label{s4}
\end{equation}%
Time does not change in this transformation since the coordinate time on the
rotating disk is given by the clock at its center and this clock is at rest
with respect to the inertial stationary disk \cite{mould}. By substituting (%
\ref{s4}) in (\ref{s3}) we obtain the metric on the rotating disk:

\begin{equation}
ds^{2}=\left( 1-\frac{\omega ^{2}r^{2}}{c^{2}}\right)
c^{2}dt^{2}-dr^{2}-r^{2}d\varphi ^{2}-2\omega r^{2}dtd\varphi -dz^{2}.
\label{s5}
\end{equation}

As light propagates along null geodesics ($ds^{2}=0$) we can calculate the
tangential coordinate velocity of light $c^{\varphi }\equiv r\left( d\varphi
/dt\right) $ from (\ref{s5}) by taking into account that $dr=0$ and $dz=0$
for light propagating on the surface of the rotating disk along its rim (of
radius $r$). First we have to determine $d\varphi /dt$. From (\ref{s5}) we
can write%
\[
r^{2}\left( \frac{d\varphi }{dt}\right) ^{2}+2\omega r^{2}\left( \frac{%
d\varphi }{dt}\right) -\left( 1-\frac{\omega ^{2}r^{2}}{c^{2}}\right)
c^{2}=0.
\]%
The solution of this quadratic equation gives two values for $d\varphi /dt$
- one in the direction in which $\varphi $ increases ($+\varphi $) (in the
direction of the rotation of the disk) and the other in the opposite
direction ($-\varphi $):

\[
\left( \frac{d\varphi }{dt}\right) ^{+\varphi }=-\omega +\frac{c}{r};\hspace{%
0.3cm}\hspace{0.5cm}\left( \frac{d\varphi }{dt}\right) ^{-\varphi }=-\omega -%
\frac{c}{r}.
\]%
Then for the tangential coordinate velocities $c^{+\varphi }$ and $%
c^{-\varphi }$ we obtain

\begin{equation}
c^{+\varphi }\equiv r\left( \frac{d\varphi }{dt}\right) ^{+\varphi }=c\left(
1-\frac{\omega r}{c}\right)  \label{c+}
\end{equation}%
and

\begin{equation}
c^{-\varphi }\equiv r\left( \frac{d\varphi }{dt}\right) ^{-\varphi
}=-c\left( 1+\frac{\omega r}{c}\right) .  \label{c-}
\end{equation}%
As seen from (\ref{c+}) and (\ref{c-}) the tangential coordinate velocities $%
c^{+\varphi }$ and $c^{-\varphi }$ are \textit{constant} for a given $r$ \
which means that (\ref{c+}) and (\ref{c-}) also represent the average
coordinate velocities of light. The coordinate speed of light propagating in
the direction of the rotation of the disk is smaller that the coordinate
speed in the opposite direction.

This fact allows an observer on the rotating disk to explain why
two light signals emitted from a point \textit{M} on the disk rim
and propagating along the rim in opposite directions will not
arrive simultaneously at \textit{M} - as the coordinate speed of
the light signal travelling against the disk rotation is greater
that the speed of the other signal it will arrive at \textit{M}
first.

The time it takes a light signal travelling along the rim of the
disk in the direction of its rotation to complete one revolution
is

\[
\Delta t^{+\varphi }=\frac{2\pi r}{c^{+\varphi }}=\frac{2\pi r}{c\left(
1-\omega r/c\right) }=\frac{2\pi r}{c-\omega r}.
\]%
The time for the completion of one revolution by the light signal
propagating in the opposite direction is:

\[
\Delta t^{-\varphi }=\frac{2\pi r}{\left\vert c^{-\varphi }\right\vert }=%
\frac{2\pi r}{c\left( 1+\omega r/c\right) }=\frac{2\pi r}{c+\omega r}.
\]%
The arrival of the two light signals at \textit{M} is separated by the time
interval:

\begin{equation}
\delta t=\Delta t^{+\varphi }-\Delta t^{-\varphi }=\frac{4\pi \omega r^{2}}{%
c^{2}-\omega ^{2}r^{2}}.  \label{s6}
\end{equation}%
The time difference (\ref{s6}) is caused by the different coordinate speeds
of light in the $+\varphi $ and $-\varphi $ directions. Here it should be
specifically stressed that $c^{+\varphi }$ and $c^{-\varphi }$ are different
from $c$ owing to the accelerated motion (rotation) of the disk. In terms of
the orbital velocity $v=\omega r$ it appears that the two tangential
coordinate velocities can be written as a function of $v$

\[
c^{+\varphi }=c\left( 1-\frac{v}{c}\right) =c-v;\hspace{0.3cm}\hspace{0.5cm}%
\hspace{0.3cm}\hspace{0.5cm}c^{-\varphi }=c\left( 1+\frac{v}{c}\right) =c+v,
\]%
which resemble the Galilean law of velocity addition. However it is
completely clear that that resemblance is misleading - due to the
centripetal (normal) acceleration $a^{N}=v^{2}/r$ the \textit{direction} of
the orbital velocity constantly changes during the rotation of the disk
which means that $c^{+\varphi }$ and $c^{-\varphi }$ depend on the normal
acceleration of the disk:

\begin{equation}
c^{+\varphi }=c\left( 1-\frac{\sqrt{a^{N}r}}{c}\right)  \label{s7}
\end{equation}%
and

\begin{equation}
c^{-\varphi }=c\left( 1+\frac{\sqrt{a^{N}r}}{c}\right) .  \label{s8}
\end{equation}%
As expected the expressions (\ref{s7}) and (\ref{s8}) are similar to the
average coordinate velocities (\ref{ca1}) and (\ref{ca2}) (for $z_{B}=0$) in
a sense that all coordinate velocities depend on acceleration, not velocity.

\section{Conclusions}

The paper revisits the question of the constancy of the speed of
light by pointing out that it has two answers - the speed of light
is constant in all inertial reference frames but when determined
in a non-inertial frame it depends on the frame's proper
acceleration. It has been shown that the complete description of
the propagation of light in non-inertial frames of reference
requires an average coordinate and an average proper velocity of
light. The need for an average coordinate velocity was
demonstrated in the case of Einstein's thought elevator experiment
- to explain the fact that
two light signals emitted from points $A$, and $C$ in Figure~1 meet at $%
B^{\prime }$, not at $B$. It was also shown that an average proper
velocity of light is implicitly used in the Shapiro time delay
effect; when such a velocity is explicitly defined it follows that
in the case of a parallel gravitational field the Shapiro effect
is not always a delay effect.

The Sagnac effect was also revisited by defining the coordinate
velocity of light in the non-inertial frame of the rotating disk.
That velocity naturally explains the fact that two light signals
emitted from a point on the rim of the rotating disk and
propagating along its rim in opposite directions do not arrive
simultaneously at the same point.

\section{Acknowledgments}

I would like to thank Mark Stuckey for his constructive and helpful comments.

\section{Appendix - Shapiro time delay}

Although it is recognized that the retardation of light (the Shapiro time
delay) is caused by the reduced speed of light in a gravitational field \cite%
[pp. 196, 197]{ohanian}, an expression for the average velocity of light has
not been derived so far. Now we shall see that the introduction of an
average proper velocity of light makes it possible for this effect to be
calculated by using this velocity. It is the average proper velocity of
light that is needed in the Shapiro time delay since the time measured in
this effect is the proper time at a given point.

We shall consider the treatment of the Shapiro time delay in \cite[Sec. 4.4]%
{ohanian}. A light (in fact, a radio) signal is emitted from the Earth (at $%
z_{1}<0$) which propagates in the gravitational field of the Sun, is
reflected by a target planet (at $z_{2}>0$), and travels back to Earth. The
path of the light signal (parallel to the $z$ axis) is approximated by a
straight line \cite[pp. 196]{ohanian}. The distance between this line and
the Sun (along the $x$ axis) is $b$. The total proper time from the emission
of the light signal to its arrival back on Earth is \cite[pp. 197, 198]%
{ohanian}:

\begin{equation}
\Delta \tau =2\left( 1-\frac{2GM_{\odot }}{c^{2}\sqrt{z_{1}^{2}+b^{2}}}%
\right) \left( \frac{z_{2}+\left| z_{1}\right| }{c}+\frac{2GM_{\odot }}{c^{3}%
}\ln \frac{\sqrt{z_{2}^{2}+b^{2}}+z_{2}}{\sqrt{z_{1}^{2}+b^{2}}-\left|
z_{1}\right| }\right) .  \label{t-ohanian}
\end{equation}

As the approximated distance between the Earth (at $z_{1}<0$) and the target
planet (at $z_{2}>0$) is $z_{2}+\left| z_{1}\right| $ we can define the
average proper velocity of a light signal travelling that distance as
determined on Earth:

\begin{eqnarray}
\bar{c}_{z_{1}z_{2}}^{g}(as\ seen\ from\ Earth) &=&\frac{z_{2}+\left\vert
z_{1}\right\vert }{\Delta \tau _{Earth}}=\frac{z_{2}+\left\vert
z_{1}\right\vert }{\Delta t}\frac{\Delta t}{\Delta \tau _{Earth}}  \nonumber
\\
&=&c_{z_{1}z_{2}}^{g}\frac{1}{\left( 1-\frac{2GM_{\odot }}{c^{2}\sqrt{%
z_{1}^{2}+b^{2}}}\right) },  \label{c-prop-schw}
\end{eqnarray}
where $c_{z_{1}z_{2}}^{g}$=$\left( z_{2}+\left\vert z_{1}\right\vert \right)
/\Delta t$ is the average coordinate velocity of light and it was taken into
account that

\[
\Delta \tau _{Earth}=\left( 1-\frac{2GM_{\odot }}{c^{2}\sqrt{z_{1}^{2}+b^{2}}%
}\right) \Delta t
\]
is the proper time as measured on Earth and obtained from the Schwarzschild
metric (the effect of the Earth's gravitational field is neglected).

\bigskip We have seen in Section 2 that the average coordinate velocity $%
c_{z_{1}z_{2}}^{g}$ can be calculated either as an average over time or over
distance, so

\[
c_{z_{1}z_{2}}^{g}=\frac{1}{z_{2}+\left\vert z_{1}\right\vert }%
\int_{z_{1}}^{z_{2}}c^{\prime }\left( z\right) dz,
\]
where

\[
c^{\prime }\left( z\right) =c\left( 1-\frac{2GM_{\odot }}{c^{2}\sqrt{%
z^{2}+b^{2}}}\right)
\]
is the coordinate velocity of light at a point in the case of the
Schwarzschild metric. Then

\begin{eqnarray*}
c_{z_{1}z_{2}}^{g} &=&\frac{c}{z_{2}+\left| z_{1}\right| }%
\int_{z_{1}}^{z_{2}}\left( 1-\frac{2GM_{\odot }}{c^{2}\sqrt{z^{2}+b^{2}}}%
\right) dz \\
&=&\frac{c}{z_{2}+\left| z_{1}\right| }\left( z_{2}+\left| z_{1}\right| -%
\frac{2GM_{\odot }}{c^{2}}\ln \frac{\sqrt{z_{2}^{2}+b^{2}}+z_{2}}{\sqrt{%
z_{1}^{2}+b^{2}}-\left| z_{1}\right| }\right) \\
&=&c\left( 1-\frac{2GM_{\odot }}{c^{2}\left( z_{2}+\left| z_{1}\right|
\right) }\ln \frac{\sqrt{z_{2}^{2}+b^{2}}+z_{2}}{\sqrt{z_{1}^{2}+b^{2}}%
-\left| z_{1}\right| }\right) .
\end{eqnarray*}
By substituting this expression for the average \textit{coordinate} velocity
of light in (\ref{c-prop-schw}) we can obtain the average \textit{proper}
velocity of light in the Schwarzschild metric:

\[
\bar{c}_{z_{1}z_{2}}^{g}(as\ seen\ from\ Earth)=\frac{c}{\left( 1-\frac{%
2GM_{\odot }}{c^{2}\sqrt{z_{1}^{2}+b^{2}}}\right) }\left( 1-\frac{2GM_{\odot
}}{c^{2}\left( z_{2}+\left\vert z_{1}\right\vert \right) }\ln \frac{\sqrt{%
z_{2}^{2}+b^{2}}+z_{2}}{\sqrt{z_{1}^{2}+b^{2}}-\left\vert z_{1}\right\vert }%
\right)
\]
or
\[
\bar{c}_{z_{1}z_{2}}^{g}(as\ seen\ from\ Earth)\approx c\left( 1+\frac{%
2GM_{\odot }}{c^{2}\sqrt{z_{1}^{2}+b^{2}}}-\frac{2GM_{\odot }}{c^{2}\left(
z_{2}+\left\vert z_{1}\right\vert \right) }\ln \frac{\sqrt{z_{2}^{2}+b^{2}}%
+z_{2}}{\sqrt{z_{1}^{2}+b^{2}}-\left\vert z_{1}\right\vert }\right) .
\]
For the total proper time
\[
\Delta \tau =\frac{2\left( z_{2}+\left\vert z_{1}\right\vert \right) }{\bar{c%
}_{z_{1}z_{2}}^{g}(as\ seen\ from\ Earth)}
\]
from the emission of the light signal to its arrival back on Earth we have

\begin{eqnarray*}
\Delta \tau &=&\frac{2\left( z_{2}+\left\vert z_{1}\right\vert \right)
\left( 1-\frac{2GM_{\odot }}{c^{2}\sqrt{z_{1}^{2}+b^{2}}}\right) }{c\left( 1-%
\frac{2GM_{\odot }}{c^{2}\left( z_{2}+\left\vert z_{1}\right\vert \right) }%
\ln \frac{\sqrt{z_{2}^{2}+b^{2}}+z_{2}}{\sqrt{z_{1}^{2}+b^{2}}-\left\vert
z_{1}\right\vert }\right) } \\
&& \\
&\approx &2\left( 1-\frac{2GM_{\odot }}{c^{2}\sqrt{z_{1}^{2}+b^{2}}}\right)
\left( \frac{z_{2}+\left\vert z_{1}\right\vert }{c}+\frac{2GM_{\odot }}{c^{3}%
}\ln \frac{\sqrt{z_{2}^{2}+b^{2}}+z_{2}}{\sqrt{z_{1}^{2}+b^{2}}-\left\vert
z_{1}\right\vert }\right)
\end{eqnarray*}%
and (\ref{t-ohanian}) is recovered. The total proper time can be also
written (to within terms proportional to $c^{-3}$) as

\[
\Delta \tau \approx 2\left( \frac{z_{2}+\left\vert z_{1}\right\vert }{c}-%
\frac{2GM_{\odot }\left( z_{2}+\left\vert z_{1}\right\vert \right) }{c^{3}%
\sqrt{z_{1}^{2}+b^{2}}}+\frac{2GM_{\odot }}{c^{3}}\ln \frac{\sqrt{%
z_{2}^{2}+b^{2}}+z_{2}}{\sqrt{z_{1}^{2}+b^{2}}-\left\vert z_{1}\right\vert }%
\right) .
\]


\begin{thebibliography}{99}
\bibitem{parallel} For the issues discussed in this paper it is sufficient
to assume that the gravitational field in an elevator at rest on the Earth's
surface is a good approximation of a parallel gravitational field.
\end{thebibliography}
\end{document}